# Losses in coated conductors under non-sinusoidal currents and magnetic fields

G. Furman , M. Spektor, V. Meerovich, V. Sokolovsky

Physics Department, Ben- Gurion University of the Negev, Beer- Sheva 84105, Israel

**Abstract**
Study of AC losses in superconducting wires and tapes is usually restricted by consideration of applied sinusoidal currents and/or magnetic fields. However, currents in electric power systems contain a wide variety of harmonics. The currents become strongly non-sinusoidal at the operation of converters, non-linear reactors, and during transient and overload conditions. Recently it has been shown that the contribution of higher harmonics to AC losses in superconducting bulk and thin film samples can be tens times larger than in normal-metal samples of the same form and the 5% harmonic can increase the losses by up to 20%. Here we report the results of the analysis of the influence of higher harmonics of the current and magnetic field on AC losses in coated conductors. Analytical expressions are obtained in the framework of the critical state model neglecting response of the normal-metal substrate and stabilization layers. Losses in the superconducting and normal-metal parts of a coated conductor are compared for various designs of the conductor. It is also shown that the 5% third current harmonic can increase the losses in the normal-metal parts by up to 90%. This increase is caused by non-linear response of superconducting layers and should be taken into account at determination of the optimal operation regimes of superconducting devices.

## I. Introduction

Currently, the main candidate for broader commercialization of high temperature superconducting (HTS) wires is the second generation (2G) $YBa_2Cu_3O_{6+x}$ (YBCO)-coated conductor. These conductors are produced by deposition of a 1-2 μm film of $YBa_2Cu_3O_{6+x}$ on ~ 50 - 100 μm thick buffered metal substrate. The superconducting film is subsequently covered by protective silver and stabilizing copper layers. The 2G superconductors are characterized by very high critical current density, up to 4 MA/cm$^2$ at 77 K, close to that in the epitaxial thin films. The implementation of superconductors in AC power applications such as transformers, motors and generators depends crucially on the success in designing conductors with low losses in time-varying magnetic environment. AC loss values determine ranges of the rated currents and magnetic fields for superconducting devices, required power of cryogenic equipment, and economical gain. Numerous investigations were devoted to AC losses in coated conductors under different conditions (see for example [1-5] and works referenced in them). However, most of works consider sinusoidal magnetic fields or/and currents. In reality, currents $I(t)$ in electric power systems contain a wide variety of harmonics and can be presented as

$$I(t) = \sum_k I_k \cos(k\omega t + \phi_k) \qquad (1)$$

where $\omega = 2\pi f$, $f$ is the frequency of the main harmonic, $I_k$ and $\phi_k$ are the amplitude and phase of the $k$-th harmonic, ($k = 0,1,2,\ldots$).

According to the requirements to power quality supported by power systems, the contribution of the higher harmonics should not exceed a few of percents. Usually higher harmonics decrease at least proportionally to $1/k$, and it is enough to take into account several first harmonics. The DC component of a current is zero because a direct current cannot be transferred by a transformer. With the accuracy of about 1%, the losses in normal conducting parts of power devices are determined by the main harmonic [6]. Currents become strongly non-sinusoidal at the operation of converters, non-linear reactors, and during transient and overload regimes. In this case all harmonics have to be taken into account at the loss estimation. Since superconductors possess a strongly non-linear current voltage characteristic, contribution of higher harmonics to the AC losses in superconducting elements can be substantial. It was shown in [6] that the contribution of higher harmonics to AC losses in superconducting bulk and thin films can be dozens of times larger than in normal-metal samples of the same form and the 5% third harmonic can increase the losses by up to 20%. Note that in normal metal, the 5% third harmonic causes the loss increase by 2% only. Moreover, the contribution of the harmonics depends on their phases: in a certain range of phases, the odd harmonics can even reduce AC losses. These peculiarities distinguish the behavior of superconducting devices from that of conventional ones.

In this paper we analyze losses in coated conductors under non-sinusoidal currents and magnetic fields.

## II. Eddy current and hysteresis losses

The total losses in a coated conductor are determined as a sum of losses in a superconductor (hysteresis losses) and in a normal-metal parts (eddy current losses). For loss calculation the coated conductors can be presented as a metal strip and a type II superconducting strip placed one on top of the other as shown in Fig. 1. The thickness of the metal strip is $d_m$ and that of the superconducting strip $d_{sc}$. Both strips are of half-width $a$ and length $l$ where $\{d_{sc}, d_m\} \ll a \ll l$, and the model of infinitely long thin strips can be applied. In the framework of the thin strip approximation the magnetic field induced by a current in the conductor is perpendicular to the surface and is independent of the $z$-coordinate inside a strip. The real current distribution through the strip thickness can be replaced by the sheet current determined as an integral of the current density through the strip thickness [7,8]. Losses in coated conductors will be calculated for cases: (i) a non-sinusoidal magnetic field is applied perpendicular to the conductor wide surface (in the $z$-direction) and (ii) non-sinusoidal current flows through the conductor in the $y$-direction without an applied magnetic field. Here losses are estimated in the framework of the critical state model neglecting the response of the normal-metal substrate and stabilization layers. For a mono-harmonic magnetic field perpendicular to a normal-metal strip the last is valid if

$$\frac{\mu_0 a d_m}{\pi \rho} \omega \ll 1, \qquad (2)$$

where $\rho$ is the resistivity of the metal. In the mathematical model of a coated conductor the substrate, protective silver and stabilizing copper layers can be presented by an effective normal-metal strip where $d_m/\rho$ is replaced by $d_{sub}/\rho_{sub} + d_{sil}/\rho_{sil} + d_{st}/\rho_{st}$ (here $\rho_{sub}$, $\rho_{sil}$, $\rho_{st}$, and $d_{sub}$, $d_{sil}$, $d_{st}$ are the resistivities and thicknesses of the substrate, protective and stabilizing layers, respectively). For a non-sinusoidal magnetic field the condition (2) has to be valid for each harmonic taken into account.

If the condition (2) is valid, the field induced by a current in the normal-metal strip may be neglected and this strip can be considered in a total field which is the sum of the applied field and field induced by a current in the superconductor.

The waveforms of non-sinusoidal currents (magnetic fields) can be divided into three types [6]: a) symmetrical case - a current monotonically increases (decreases) from the minimum (maximum) to the maximum (minimum), and the maximum equals the module of the minimum; b) asymmetrical case - a current monotonically changes like in the first case but the maximum does not equal the module of the minimum; c) a current change is non-monotonic. The type of a waveform depends on the number of harmonics, their frequencies and phases. We will consider the cases where a current (magnetic field) monotonically changes.

II.a. *Coated conductor in a perpendicular magnetic field*

A non-sinusoidal magnetic field $H_e(t)$ is applied perpendicular to the surface of a coated conductor. It is known that if a external magnetic field monotonically increases, in the framework of the critical state model the AC losses in a superconductor are determined only by the extreme values and do not depend on the waveform. Using approaches developed in [7] we obtained the following expressions for hysteresis losses per length unit of a thin strip for the asymmetrical case b):

$$P_h = 4\mu_0 a^2 f J_c H_f \Delta h \, g(\Delta h), \tag{3}$$

where $\Delta h = (H_{max} - H_{min})/2H_f$, $H_{max}$ and $H_{min}$ are the maximum and minimum of the magnetic field, respectively, $H_f = J_c/\pi$, $J_c = j_c d_{sc}$, $j_c$ is the critical current density, $g(t) = \frac{2}{t} \ln \cosh t - \tanh t$.

The eddy current loss power $p_{ed}$ per unit of volume in an infinity long thin strip is a function of $x$ and time $t$:

$$p_{ed} = j_{y,ed}(x,t) E_y(x,t), \tag{4}$$

where $E_y$ and $j_{y,ed}$ are the electric field and current density inside the metal pointing in the y-direction and related according to Ohm's law:

$$E_y = \rho j_{y,ed}. \tag{5}$$

The electric field $E_y$ in the metal strip is caused by the time variation of the magnetic field

$$E_y(x,t) = -\mu_0 \int_0^x \frac{\partial H_z(u,t)}{\partial t} du, \tag{6}$$

here, $H_z$ is the component of the magnetic field in the z-direction which is the sum of the applied field $H$ and the z-component of the magnetic field generated by the induced currents in the superconducting layer at $z = 0$. The eddy losses per length unit are:

$$P_{ed} = \mu_0^2 \frac{\omega}{2\pi} \frac{d_m}{\rho} \int_0^{2\pi/\omega} dt \int_{-a}^{a} dx \left[ \frac{\partial H_z(x,t)}{\partial t} \right]^2. \tag{7}$$

Following [5,7] for the external field $H_e$ increased from zero (from the virgin state) the penetrated field is

$$H_z(x, H_e, J_c) = \begin{cases} H_f \operatorname{arctanh} \dfrac{\sqrt{x^2 - b^2}}{c|x|}, & b < |x| < a \\ 0 & |x| < b \end{cases}, \quad (8)$$

where $b = \dfrac{a}{\cosh(H_e / H_f)}$, $c = \tanh(H_e / H_f)$.

Let us assume that $H_{max} \geq |H_{min}|$. In this case the penetrated field at $H_e = H_{max}$ is given by $H_z(x, H_{max}, J_c)$. At the decrease of the external field from the maximum till the minimum the penetration field is determined as $H_z = H_z(x, H_{max}, J_c) - H_z(x, H_{max} - H_e, 2J_c)$. At the $H_e$ increase from the minimum to the maximum the penetration field is presented as $H_z = H_z\big|_{H_e = H_{min}} + H_z(x, H_{min} - H_e, 2J_c)$. Note that only the second terms give a contribution into eddy losses. Eddy power losses per length unit can be sought as

$$P_{ed} = \frac{\mu_0^2 d_m a^3 \omega^2}{3\pi \rho} H_1^2 F_h \quad (9)$$

where

$$F_h = \frac{1}{\omega \pi} \left\{ \int_{t_1}^{t_2} (\partial h_e)^2 \left[ 1 - \frac{3}{\cosh^2(Y)} + \frac{2}{\cosh^3(Y)} \right] dt + \int_{t_2}^{2\pi/\omega + t_1} (\partial h_e)^2 \left[ 1 - \frac{3}{\cosh^2(X)} + \frac{2}{\cosh^3(X)} \right] dt \right\} \quad (10)$$

where $\partial h_e = \dfrac{1}{H_1} \dfrac{\partial H_e(t)}{\partial t}$, $Y = \dfrac{H_{max} - H_e}{2H_f}$, $X = \dfrac{H_{min} - H_e}{2H_f}$, $H_1$ is the amplitude of the main harmonic, $t_1$, is $t_2$ are the points at which the external field reaches its maximum and minimum, respectively. We obtain the following formula for total power losses per length unit in a coated conductor:

$$\frac{P_{Htot}}{(\mu_0^2 d_m a^3 / 3\rho) H_1^2 \omega^2} = p_H = F_h + c_h Q_h, \quad (11)$$

where $Q_h = 6 \dfrac{\Delta h}{h_1^2} g(\Delta h)$, $c_h = \dfrac{\pi \rho}{\mu_0 a d_m \omega}$, $h_1 = \dfrac{H_1}{H_f}$.

Here losses are normalized to the eddy losses caused by the uniform magnetic field with the amplitude and frequency equaled to those of the main harmonic when a superconductor is absent.

II.b. *Coated conductor with a transport current*

Let us consider losses caused by a non-sinusoidal transport current $I(t)$ flowing in the y-direction without an external magnetic field. We study the case when the current maximum $I_{max}$ and module of the current minimum $I_{min}$ are less than the critical value $I_c = 2aJ_c$. It is assumed that the magnetic field generated by an induced current in the metal is negligible in comparison with the field from the current in the superconductor. Losses in the superconductor are given by [6]:

$$P_{hys} = \frac{\mu_0 \omega I_c^2}{2\pi^2} q\left( \frac{I_{max} - I_{min}}{2I_c} \right), \quad (12)$$

where $q(x) = (1-x)\ln(1-x) + (1+x)\ln(1+x) - x^2$.

To calculate the eddy current loss power $P_{ied}$ per length unit, Eq. (7) is used where, at a decrease of the current from its maximum till the minimum, the magnetic field is determined as [7]:

$$H_z(x,t) = \tilde{H}_z(x, I_{max}, I_c) - \tilde{H}_z(x, I_{max} - I(t), 2I_c) \tag{13}$$

where $\tilde{H}_z(x,I,I_c) = \begin{cases} 0, & |x| < b_i \\ 2H_f \dfrac{x}{|x|} \operatorname{arctanh}\left[\dfrac{x^2 - b_i^2}{a^2 - b_i^2}\right]^{1/2}, & b_i < |x| < a \end{cases}$, and $b_i = a\sqrt{1 - I^2/I_c^2}$.

Here as in the case of the external field, we assume that $I_{max} \geq |I_{min}|$ and, hence, the penetrated field at $I = I_{max}$ is given by $H_z(x, I_{max}, J_c)$. At the decrease of the transport current from the maximum till the minimum, the penetration field is determined as $H_z = H_z(x, I_{max}, J_c) - H_z(x, I_{max}-I(t), 2J_c)$. At the current increase from the minimum to the maximum, the penetrated field is presented as $H_z = H_z\big|_{I=I_{min}} + H_z(x, I_{min} - I(t), 2J_c)$ and, again, only the second term gives a contribution into eddy current losses. The total power losses per length unit in a coated conductor are:

$$\frac{P_{ltot}}{\left(\mu_0^2 d_m a^3 / 2\rho\pi^3\right) I_1^2 \omega^2} = p_I = F_i + c_h Q_i(\Delta i), \tag{14}$$

where $\Delta i = \dfrac{I_{max} - I_{min}}{2I_c}$, $Q_i(x) = q(x)/i_1^2$, $i_1 = I_1/I_c$,

$$F_i = \frac{1}{\omega}\left\{\int_{t_1}^{t_2} \partial i(t)^2 G(b_i) dt + \int_{t_2}^{2\pi/\omega + t_1} \partial i(t)^2 G(\tilde{b}_i) dt\right\},$$

$$G(x) = 1 - x - \sqrt{1-x^2}\operatorname{Log}\left(\frac{1+\sqrt{1-x^2}}{x}\right) + \frac{1}{2}\left[\operatorname{Log}\left(\frac{1+\sqrt{1-x^2}}{x}\right)\right]^2,$$

$$b_i = \sqrt{1 - \left(\frac{I_{max} - I(t)}{2I_c}\right)^2}, \quad \tilde{b}_i = \sqrt{1 - \left(\frac{I_{min} - I(t)}{2I_c}\right)^2}, \quad \partial i(t) = \frac{1}{I_1}\frac{dI(t)}{dt}.$$

Note, that functions $F_h$ and $F_i$ characterizing the eddy losses as well as functions $Q_h$ and $Q_i$ corresponding to losses in superconductor are independent of frequency.

### III. Results and discussion

In the case of a sinusoidal waveform ($H_e(t) = H_1 \sin(\omega t)$ or $I(t) = I_1 \sin(\omega t)$), the dependence of the functions $F_h$, $F_i$, $G_h$ and $G_i$ on the magnetic field and current has been analyzed in [5]. We will consider two first harmonics $k = 2$ and $k = 3$. The second harmonic exists in a two-phase network, the third ($k = 3$) harmonic – in a three-phase systems. The expressions for losses were obtained under the assumption that the external magnetic field (transport current) monotonically changes between its extrema. For $I(t) = I_1\sin(\omega t) + I_k\sin(k\omega t)$, the monotonic behavior is observed at $I_k/I_1 < 1/k^2$. The functions $F_h$, $G_h$ and $G_i$, $F_i$ are shown in Figs. 2 and 3, respectively.

The influence of higher harmonics on losses in a superconducting strip in perpendicular magnetic field has been analyzed in [6], where it was shown that the hysteresis losses are independent of frequency of the harmonics. These losses depend only on the maximum and minimum of the current or magnetic field. Therefore, the AC loss calculation accounting the main harmonic only leads to an error determined by the difference $\Delta H = (H_{max} - H_{min}) - 2H_1$, for example. To a first approximation, the losses can be estimated as

$$P \approx P_1(1 + K \Delta H / 2H_1),  \quad (15)$$

where $P_1$ is the AC losses caused by the main harmonic, $K = 4$ for a strip at $H_{max} - H_{min} \ll H_f$ and $K = 1$ at high magnetic fields. A relatively low difference $\Delta H/2H_1$ leads to a noticeable increase of the AC losses (Fig. 4a). For example, at $\Delta H/2H_1 = 0.05$ the losses increase by up to 20% at low field and up to 5% at high field. At the same time, in the normal metal, the 5% second harmonic causes the loss increase by 1%. The difference $\Delta H$ is determined not only by amplitudes of the harmonics but also their phases. For $k = 2$, the maximum of $\Delta H$ is achieved at $\phi_2 = 0$ and increases as about $2.4 H_2^{1.8}$, for $k = 3$ this maximum equals $2H_3$ at $\phi_3 = \pi$. Our calculation shows that the eddy losses also are highest at these phases. At design of superconducting power devices, one should calculate AC losses for the worst case when the losses are maximal, therefore here we will analyze the case of the third harmonic with $\phi_3 = \pi$. Consideration of the second harmonic give gives qualitatively close results.

The relative contribution $\delta P = (P - P_1)/P_1$ of higher harmonics to losses is maximal at low magnetic field and decreases with the field increase (Fig. 4). Another dependence is obtained for a superconducting strip with current (Fig. 5a): at $I_{max} - I_{min} \ll I_c$ losses are determined by the expression similar to (15) with $K = 4$, where the magnetic field values are replaced by the corresponding currents. At $\Delta I/2I_1 = 0.05$ the losses increase also by up to 20%. However, in contrast to the case of magnetic field, the relative contribution increases with the current and achieves about 35% at $I \approx I_c$. Qualitatively similar behavior is observed for losses in the normal-metal parts. The relative contribution increases with a current and decreases when an external field increases (Figs. 4b and 5b). However, the relative contribution can be more substantial. For example, the 5%-third harmonic causes increase of eddy losses by about 35% at low currents. At a high field when screen properties of a superconductor can be neglected, eddy losses are determined as for a normal-metal strip in the uniform external non-sinusoidal field and the relative contribution of the $k$-th harmonic to the losses is given by $\delta P = k^2 H_k^2 / H_1^2$. Higher harmonics of the current cause a greater increase of losses than that caused by the magnetic field harmonics. Existence of the 5%-third harmonic leads to the 50% loss increase at low currents and to about 90% at $I \approx I_c$.

The total losses in a coated conductor are determined by the sum of the losses in the superconducting and normal-metal strips. The relative contributions of these parts are determined by the ratio $F_h$ and $G_h$ in case of the magnetic field and by the ratio $F_i$ and $G_i$ for the current as well as by the constant $c_h$. At $h_1 \ll 1$ both functions, $F_h$ and $G_h$, are well fitted by power laws with exponents 4 and 2, respectively. At $h_1 \gg 1$ $G_h$ decreases proportionally to $1/h_1$ while $F_h$ tends to the limiting value $1 + (kH_k/H_1)^2$. Thus, the relative contribution of eddy current losses in the total losses increases with the applied field (Fig. 2). In spite of the fact that both functions $F_i$ and $G_i$ increase (Fig. 3) with current, the relative contribution of losses in a normal-metal strip also increase. At $i_1 = 0.2$ the ratio $F_i/G_i = 0.02$; at $i_1 = 0.9$ this ratio is about 0.6.

Let us estimate the parameter $c_h$ which is the same in the expressions for the total losses in both cases and depends on frequency of the main harmonic and characteristics of the normal-metal strip (resestivity, thickness, and width) only. For a coated conductor without protective silver layer and stabilizer [5] the parameter $c_h$ is evaluated as ($10^6$ s-

$^1)/f$, for a conductor with the silver layer with thickness of ~2 μk [9] $c_h \approx (8 \cdot 10^4 \text{ s}^{-1})/f$, for a well stabilized coated conductor with cooper stabilizer thickness of 100 μk [10] the parameter $c_h$ is evaluated as $(1.6 \cdot 10^3 \text{ s}^{-1})/f$. The estimations were done for the coated conductor with the width of 1 cm, the resistivity of a hastelloy-C substrate was taken $1.24 \cdot 10^{-6}$ Ω m and resistivity of silver and cooper – $2 \cdot 10^{-9}$ Ω m at 77 K. One can see that the losses in a superconductor dominate in the two first conductor types till high frequencies of the order of 1 MHz. The losses in the normal-metal parts of a well stabilized coated conductor can be comparable or even dominate at electrotechnical frequencies ~1 kHz (the range of frequency is used in special electric systems such as airplanes, ships, etc.). At 50-60 Hz losses in the normal-metal parts will dominate in magnetic field $h_1 > 30$, e.i. when $\mu_0 H_1 > 0.1$ T at the typical value of $J_c = 10^4$ A/cm. This magnetic field is much less than the working fields in many power devices, for example, a working field of a generator is the order of 1 T.

### IV. Conclusion

Our results show that higher harmonics in coated conductors can substantially change the amount of power losses, especially, in the conductors with a transport current. While, in the normal metal, the 5% third current harmonic causes the loss increase by about 2%, in the superconducting part of a coated conductor, this increase can achieve 50% while in the normal-metal parts – 90%. These peculiarities distinguish the behavior of coated conductors from that conventional ones and show necessity of taking into account higher harmonics with even low amplitudes at the loss and thermal state analysis of the coated conductors.

### Acknowledgements

This research was supported by a Grant from the Ministry of Science, Culture & Sport of Israel and the Russian Foundation for Basic Research, the Russian Federation.

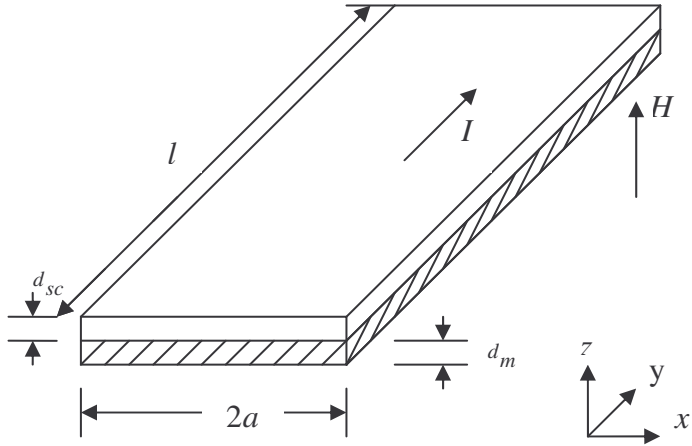

Fig. 1 Sketch of the coated conductor with a transport current $I$ in a magnetic field $H$.

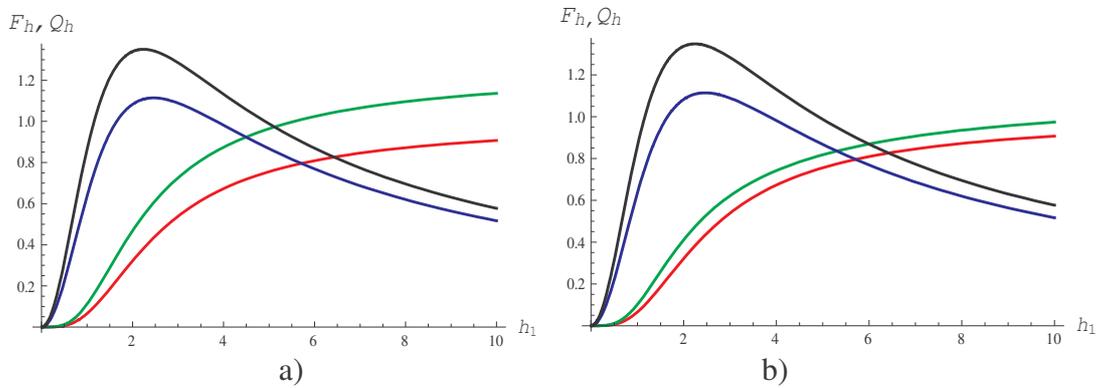

Fig. 2 Dependences of the functions $F_h$ and $Q_h$ on amplitudes of two first harmonics: a) the main and second harmonics at $\phi_2 = 0$ (blue, $Q_h$, and red, $F_h$, lines at $H_2 = 0$; black, $Q_h$, and green, $F_h$, lines $H_2 = 0.25\, H_1$) and b) main and third harmonics at $\phi_3 = \pi$ (blue, $Q_h$, and red, $F_h$, lines at $H_3 = 0$; black, $Q_h$, and green, $F_h$, lines at $H_3 = 0.1\, H_1$).

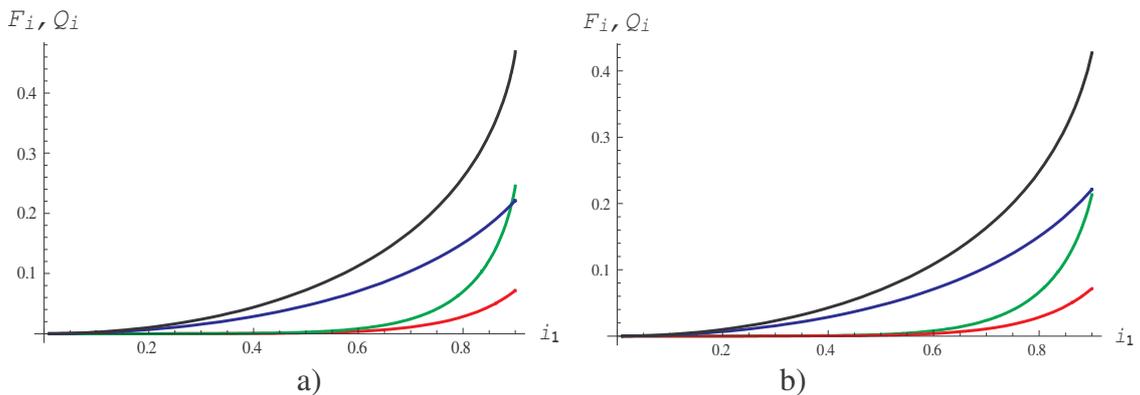

Fig. 3 Dependences of the functions $F_i$ and $Q_i$ on amplitudes of two first harmonics: a) the main and second harmonics at $\phi_2 = 0$ (blue, $Q_i$, and red, $F_i$, lines at $I_2 = 0$, black, $Q_i$, and green, $F_i$, lines at $I_2 = 0.25\, I_1$) and b) main and third harmonics at $\phi_3 = \pi$ (blue, $Q_i$, and red, $F_i$, lines at $I_3 = 0$; black, $Q_i$, and green, $F_i$, lines at $I_3 = 0.1\, I_1$).

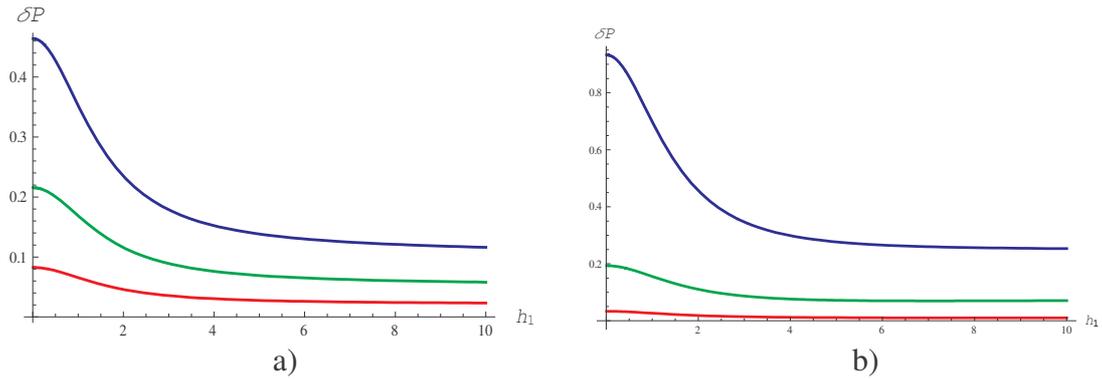

Fig. 4 Relative contribution of the third magnetic field harmonic to losses in a superconductor (a) and normal-metal parts (b): blue line - $H_3 = 0.1\, H_1$; green – $H_3 = 0.05\, H_1$, red line – $H_3 = 0.02\, H_1$.

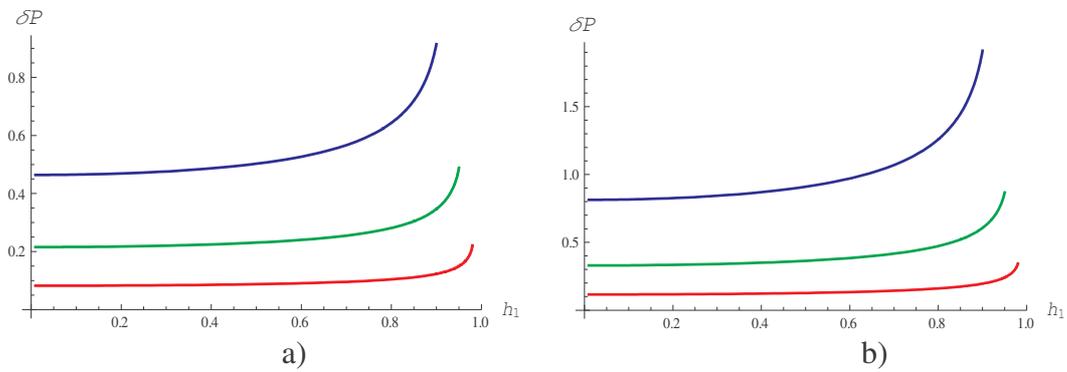

Fig. 5 Relative contribution of the third current harmonic to losses in a superconductor (a) and normal-metal parts (b): blue line - $I_3 = 0.1 I_1$, green - $I_3 = 0.05 I_1$, red line - $I_3 = 0.02 I_1$.